\documentclass[useAMS,usenatbib]{mn2e}
\usepackage{graphicx}
\usepackage{times}

\newcommand{\MSUN}{\mbox{$\mathrm{M_{\odot}}$}}

\title[New eclipsing PCEBs from CRTS]{Fourteen new eclipsing white dwarf plus
  main-sequence binaries from the SDSS and Catalina surveys}

\author[S. G. Parsons et al.]{S.~G.~Parsons$^{1}$\thanks{steven.parsons@uv.cl},
C.~Agurto-Gangas$^{1}$,
B.~T.~G{\"a}nsicke$^{2}$,
A.~Rebassa-Mansergas$^{3}$
\newauthor
M.~R.~Schreiber$^{1,4}$,
T.~R.~Marsh$^{2}$,
V.~S.~Dhillon$^{5}$,
S.~P.~Littlefair$^{5}$,
A.~J.~Drake$^{6}$,
\newauthor
M.~C.~P.~Bours$^{2}$,
E.~Breedt$^{2}$,
C.~M.~Copperwheat$^{7}$,
L.~K.~Hardy$^{5}$,
C.~Buisset$^{8}$,
\newauthor
P.~Prasit$^{8}$
and J.~J.~Ren$^{9}$
\\
$^{1}$ Departamento de F{\'i}sica y Astronom{\'i}a, Universidad de
Valpara{\'i}so, Avenida Gran Bretana 1111, Valpara{\'i}so, Chile\\
$^{2}$ Department of Physics, University of Warwick, Coventry CV4 7AL, UK\\
$^{3}$ Kavli Institute for Astronomy and Astrophysics, Peking University, Beijing 100871, China\\
$^{4}$ Millenium Nucleus "Protoplanetary Disks in ALMA Early Science",
Universidad de Valparaiso, Valparaiso 2360102, Chile\\
$^{5}$ Department of Physics and Astronomy, University of Sheffield,
Sheffield, S3 7RH, UK\\
$^{6}$ California Institute of Technology, 1200 E. California Blvd, CA 91225,
USA\\
$^{7}$ Astrophysics Research Institute, Liverpool John Moores University, Liverpool L3 5RF, UK\\
$^{8}$ National Astronomical Research Institute of Thailand, 191 Siriphanich
Building, Huay Kaew Road, Chiang Mai 50200, Thailand\\
$^{9}$Department of Astronomy, Peking University, Beijing 100871, P.\,R.\,China}
\begin{document}
\input{references.cls}
\date{Accepted 2015 February 18.  Received 2015 February 17; in original form 2015 February 2}

\pagerange{\pageref{firstpage}--\pageref{lastpage}} \pubyear{2015}

\maketitle

\label{firstpage}

\begin{abstract}

We report on the search for new eclipsing white dwarf plus main-sequence
(WDMS) binaries in 
the light curves of the Catalina surveys. We use a colour selected list of
almost 2000 candidate WDMS systems from the Sloan Digital Sky Survey,
specifically designed to identify WDMS systems with cool white dwarfs and/or
early M type main-sequence stars. We identify
a total of 17 eclipsing systems, 14 of which are new discoveries. We also
find 3 candidate eclipsing systems, 2 main-sequence eclipsing binaries and
22 non-eclipsing close binaries. Our newly discovered systems generally have
optical fluxes dominated by the main-sequence components, which have earlier
spectral types than the majority of previously discovered eclipsing systems.
We find a large number of ellipsoidally variable binaries with similar
periods, near 4 hours, and spectral types M2--3, which are very close to
Roche-lobe filling. We also find that the fraction of eclipsing systems is
lower than found in previous studies and likely reflects a lower close binary
fraction among WDMS binaries with early M-type main-sequence stars due to
their enhanced angular momentum loss compared to fully convective late M type
stars, hence causing them to become cataclysmic variables quicker and
disappear from the WDMS sample. Our systems bring the total number of known
detached, eclipsing WDMS binaries to 71.

\end{abstract}


\section{Introduction}

The majority of main-sequence star binaries ($\sim75\%$) are
sufficiently well separated that their stellar components can evolve
independently of each other and effectively as single stars. The remaining
$\sim25\%$ have separations small enough that when the more massive member of
the binary ascends the giant branch, the two stars will interact with each
other causing mass transfer via Roche-lobe overflow \citep{willems04}. This
interaction often leads to a common envelope phase, during which the two stars
orbit within a single envelope of material, quickly losing angular momentum
and spiraling towards each other \citep{webbink84,ivanova13}. The result of
this phase is a compact binary with a period of hours to days, comprised of
the core of the initially more massive star and its main-sequence companion,
known as a post common envelope binary (PCEB). 

\begin{table*}
 \centering
  \caption{Journal of photometric observations. The eclipse of the white dwarf occurs at phase 0, 1 etc.}
  \label{tab:photobslog}
  \begin{tabular}{@{}lcccccc@{}}
  \hline
  Target           & Date at      & Telescope     & Filter(s) & Start & Orbital & Exposure \\
                   & start of run & \& instrument &           & (UT)  & phase   & time (s) \\
  \hline
  SDSS\,J002412.87+174531.4 & 2013/07/09   & LT+RISE       & V+R       & 04:19 & 0.92 -- 1.06 & 10.0     \\
  SDSS\,J042012.78+333739.7 & 2014/03/05   & WHT+ULTRACAM  & $u'g'r'$  & 20:15 & 0.79 -- 1.06 & 5.2      \\
  SDSS\,J101356.32+272410.6 & 2012/12/19   & LT+RISE       & V+R       & 04:04 & 0.92 -- 1.14 & 6.0      \\
  SDSS\,J112308.40-115559.3 & 2014/01/28   & TNT+ULTRASPEC & KG5       & 17:19 & 0.92 -- 1.04 & 7.8      \\
  SDSS\,J141150.74+211750.0 & 2013/05/12   & LT+RISE       & V+R       & 01:56 & 0.96 -- 1.08 & 10.0     \\
  SDSS\,J141724.36+080112.0 & 2014/06/18   & INT+WFC       & $g$       & 22:59 & 0.91 -- 1.09 & 20.0     \\
  SDSS\,J142427.69+112457.9 & 2013/04/22   & WHT+ULTRACAM  & $u'g'r'$  & 23:29 & 0.92 -- 1.03 & 2.2      \\
  SDSS\,J154057.27+370543.4 & 2014/03/27   & LT+RISE       & V+R       & 04:00 & 0.88 -- 1.11 & 10.0     \\
  SDSS\,J164235.97-063439.7 & 2014/04/22   & LT+RISE       & V+R       & 03:53 & 0.91 -- 1.08 & 10.0     \\
  SDSS\,J220504.50-062248.6 & 2012/09/10   & WHT+ULTRACAM  & $u'g'r'$  & 23:00 & 0.80 -- 1.07 & 3.1      \\
  SDSS\,J230627.54-055533.2 & 2015/01/02   & TNT+ULTRASPEC & KG5       & 12:16 & 0.85 -- 1.09 & 4.1      \\
  \hline
  \end{tabular}
\end{table*}

The most abundant type of PCEB are those containing a white dwarf with a
low-mass main-sequence star companion. In recent years the number of these
systems known has rapidly increased, thanks mainly to the Sloan Digital Sky
Survey (SDSS; \citealt{york00,adelman08,abazajian09}). The number of white
dwarf plus main-sequence binaries (WDMS) spectroscopically observed within
SDSS has reached 2316 as of data release 8 \citep{rebassa13}, with an
additional 227 recently discovered in data release 9 \citep{li14}. From this
large sample more than 200 close PCEB systems have been identified
\citep{nebot11,parsons13css}. These systems have been used to investigate many
aspects of close binary evolution, such as disrupted magnetic braking
\citep{schreiber10}, the common envelope efficiency \citep{zorotovic10}, the
origin of low-mass white dwarfs \citep{rebassa11} and activity in rapidly
rotating M dwarfs \citep{rebassa13act}. Additional surveys are also adding to
these numbers (e.g. the LAMOST surveys, \citealt{ren14}). 

However, one limitation of the SDSS WDMS sample is that it is biased towards
systems containing M dwarf stars with late spectral types and relatively hot
white dwarfs. This is because the white dwarfs in these systems contribute a
relatively large fraction of the optical flux, thus making their spectral
features visible (hence easier to detect) and altering the colours of the
objects sufficiently to move them well off the main-sequence, hence making
these types of WDMS systems easier to identify. Moreover, the similar colours
of WDMS of this type to quasars, which were specifically targeted within the SDSS
spectroscopic survey, meant that these systems are overrepresented within the
spectroscopic sample. WDMS binaries containing early-M or more massive
main-sequence components and cooler white dwarfs generally have optical fluxes
dominated by the main-sequence star making their identification as
WDMS systems difficult, although 251 systems of this type were identified within the
Sloan Extension for Galactic Understanding and Exploration (SEGUE) survey
\citep{rebassa12}.

Eclipsing PCEBs are particularly useful since the eclipse
allows us to directly measure the physical parameters of the binary virtually
independent of model atmosphere calculations and hence test mass-radius
relationships
\citep{parsons10nn,parsons12gs,parsons12uc,pyrzas12,littlefair14}. Moreover,
the sharp white dwarf eclipse features lead to extremely precise timing
measurements and have led to the discovery of quasi-sinusoidal variations in
the eclipse arrival times of many of these systems
\citep{parsons10omc,backhaus12} which, in some cases, have been interpreted as
the gravitational influence of circumbinary planets
\citep{beuermann10,beuermann12ii,beuermann12i,marsh14}. 

Recently \citet{rebassa13} used a photometric selection criteria specifically
designed to target WDMS systems in the SDSS dominated by the contribution from
the main-sequence star, hence systems containing early type M dwarfs and/or
cool white dwarfs. In this paper we combine this photometric catalogue with
data from the Catalina Sky Survey (CSS) and Catalina Real Time Transient
Survey (CRTS; \citealt{drake09}) in order to search for new eclipsing
systems. We also provide a full list of known detached, eclipsing white dwarf
plus main-sequence / brown dwarf binaries published to date. 

\section{Sample selection and data reduction}

We use the list of 3419 photometrically selected white dwarf plus
main-sequence binary candidates from \citet{rebassa13}. These were selected
based on a combination of optical (SDSS) and infrared (UKIRT, 2MASS and WISE)
colours. The catalogue also contains 47 spectroscopically confirmed white
dwarf plus main-sequence binaries from SDSS DR 8. We select those systems with
magnitudes of $g<19$ and CSS coverage, resulting in a total sample of 2060
systems. 

We re-reduced the raw CSS data ourselves using the method outlined in
\citet{parsons13css}, which allowed us to more easily identify deeply
eclipsing systems as well as remove any contaminated exposures. Some of our
objects were completely blended with nearby stars and so we remove these from
our sample. In total we had 1958 objects with good CSS photometry. 

\section{Follow-up data}

We obtained follow-up photometry and spectroscopy for a number of our
systems. In this section we outline those observations and their
reduction. Unfortunately, due to limited observing time, we have been
  unable to obtain follow-up high-speed photometry for 3 of the new eclipsing
  systems. A full summary of the photometric observations is given in
Table~\ref{tab:photobslog}. 

\subsection{William Herschel Telescope + ULTRACAM photometry}

We observed three of our new eclipsing systems with the high speed
frame-transfer camera ULTRACAM \citep{dhillon07} mounted as a visitor
instrument on the 4.2-m William Herschel Telescope (WHT) on La Palma. The
observations targeted the eclipse of the white dwarf. ULTRACAM uses a triple
beam setup allowing one to obtain data in three separate bands simultaneously
with a deadtime between frames of only 24 milliseconds. For all of our
observations we used ULTRACAM equipped with $u'$, $g'$ and $r'$ filters. 

All of these data were reduced using the ULTRACAM pipeline software
\citep{dhillon07}. Debiassing, flatfielding and sky background subtraction
were performed in the standard way. The source flux was determined with
aperture photometry using a variable aperture, whereby the radius of the
aperture is scaled according to the full width at half maximum
(FWHM). Variations in observing conditions were accounted for by determining
the flux relative to a comparison star in the field of view. 

\subsection{Liverpool Telescope + RISE photometry}

We used the high speed camera RISE \citep{steele08} on the Liverpool Telescope
(LT) to observe the eclipses of five of our systems. RISE is a frame transfer
CCD camera with a single wide-band V+R filter and negligible deadtime between
frames. The raw data are automatically run through a pipeline that debiases,
removes a scaled dark frame and flat-fields the data. We used the ULTRACAM
pipeline to perform aperture photometry on the RISE data in the same way as
described in the previous section. 

\subsection{Thai National Telescope + ULTRASPEC photometry}

Two of our eclipsing systems were observed with the high-speed ULTRASPEC
camera \citep{dhillon14} mounted on the 2.4-m Thai National Telescope (TNT),
located on Doi Inthanon, Thailand. ULTRASPEC has a frame transfer EMCCD with a
deadtime between exposures of 15 milliseconds. Our observations were designed
to cover the white dwarf eclipses. We used the Schott KG5 filter, which is a
broad filter with a central wavelength of 5075{\AA} and FWHM of 3605{\AA} and
hence covers the $u'$, $g'$ and $r'$ bandpasses. We again used the ULTRACAM
pipeline to reduce the data.

\subsection{Isaac Newton Telescope + WFC photometry}

We observed one of our new eclipsing systems with the Wide Field Camera (WFC)
mounted at the prime focus of the 2.5-m Isaac Newton Telescope (INT) on La
Palma. The observations were performed with the Sloan-Gunn $g$ filter and we
windowed the detector in order to reduce the read-out time to $\sim$2
seconds. Once again the ULTRACAM pipeline was used to reduce the data. 

\begin{table}
 \centering
  \caption{Journal of NAOC+BFOSC spectroscopic observations.}
  \label{tab:specobslog}
  \begin{tabular}{@{}lccc@{}}
  \hline
  Target           & Date at      & UT             & Exposure    \\
                   & start of run & (mid-exposure) & time (mins) \\
  \hline
  SDSS\,J0031+2634 & 2014/08/06   & 20:04 & 25 \\
  SDSS\,J0046-0339 & 2014/08/06   & 19:33 & 25 \\
  SDSS\,J0054+1429 & 2014/08/06   & 19:07 & 35 \\
  SDSS\,J0249+0713 & 2014/08/04   & 20:08 & 20 \\
  SDSS\,J1411+2117 & 2014/08/07   & 13:19 & 25 \\
  SDSS\,J1506+2152 & 2014/08/06   & 13:31 & 30 \\
  SDSS\,J1540+3705 & 2014/08/06   & 15:05 & 25 \\
  SDSS\,J2053+0015 & 2014/08/04   & 16:31 & 20 \\
  SDSS\,J2122+1039 & 2014/08/06   & 15:56 & 35 \\
  SDSS\,J2148+1927 & 2014/08/06   & 14:34 & 40 \\
  SDSS\,J2205-0622 & 2014/08/04   & 19:47 & 45 \\
  SDSS\,J2229+1853 & 2014/08/04   & 17:51 & 30 \\
  SDSS\,J2241+2536 & 2014/08/04   & 17:11 & 25 \\
  SDSS\,J2256+1822 & 2014/08/04   & 18:50 & 40 \\
  SDSS\,J2327+2119 & 2014/08/06   & 17:01 & 40 \\
  SDSS\,J2339+1157 & 2014/08/06   & 18:20 & 30 \\
  SDSS\,J2340+1855 & 2014/08/06   & 17:47 & 35 \\
  \hline
  \end{tabular}
\end{table}

\subsection{NAOC Xinglong + BFOSC spectroscopy}

\begin{figure}
\begin{center}
 \includegraphics[width=0.98\columnwidth]{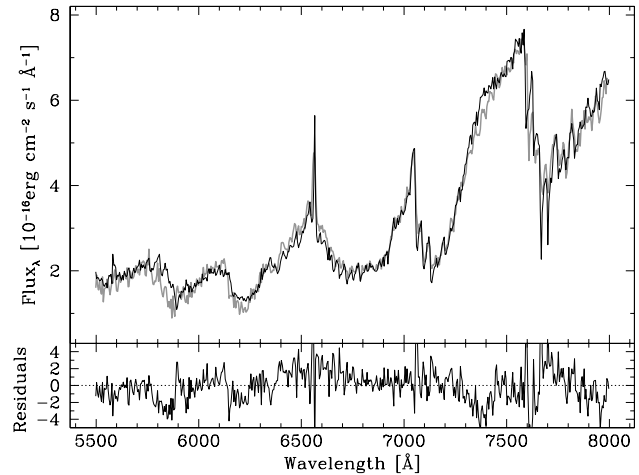}
 \caption{BFOSC Spectrum of one of our newly identified PCEB systems
   (SDSS\,J2241+2536, gray line) with the best fit template spectrum
   over-plotted (black line). The lower panel shows the residuals of the fit.}
 \label{fig:specfit}
 \end{center}
\end{figure}

Low-resolution spectroscopic observations of 17 targets showing variation in
their light curves were performed with the 2.16-m optical telescope at the
Xinglong Station of the National Astronomical Observatories, Chinese Academy
of Sciences (NAOC), using the BAO Faint Object Spectrograph and Camera
(BFOSC). The observations were performed in August 2014 and we covered as
  many close systems (both eclipsing and non-eclipsing) as were visible at the
  time. We used the low resolution grism-G5 with a dispersion degree of 1.99
\AA/mm, a spectral resolution of 2.98{\AA} and a long-slit of width 1.8''. The
spectra cover the wavelength range of 5200{\AA}--10120{\AA}. The typical seeing
varied from 1.8'' to 3.0''. A summary of these observations is given in
Table~\ref{tab:specobslog}. 

The spectroscopic data were reduced following the grating spectroscopy
procedures provided in the {\sc pamela} and {\sc molly}\footnote{{\sc pamela}
  and {\sc molly} are available from
  www2.warwick.ac.uk/fac/sci/physics/research/astro/people/marsh/software}
packages. The spectra were first bias subtracted and flat-fielded, then the
object spectra were optimally extracted. The wavelength calibration was
performed with Fe/Ar lamps exposed at the beginning and the end of each
night. The observations of spectral standard stars were used to flux calibrate
and remove the telluric features from our spectra. 

\begin{figure*}
\begin{center}
 \includegraphics[width=0.99\textwidth]{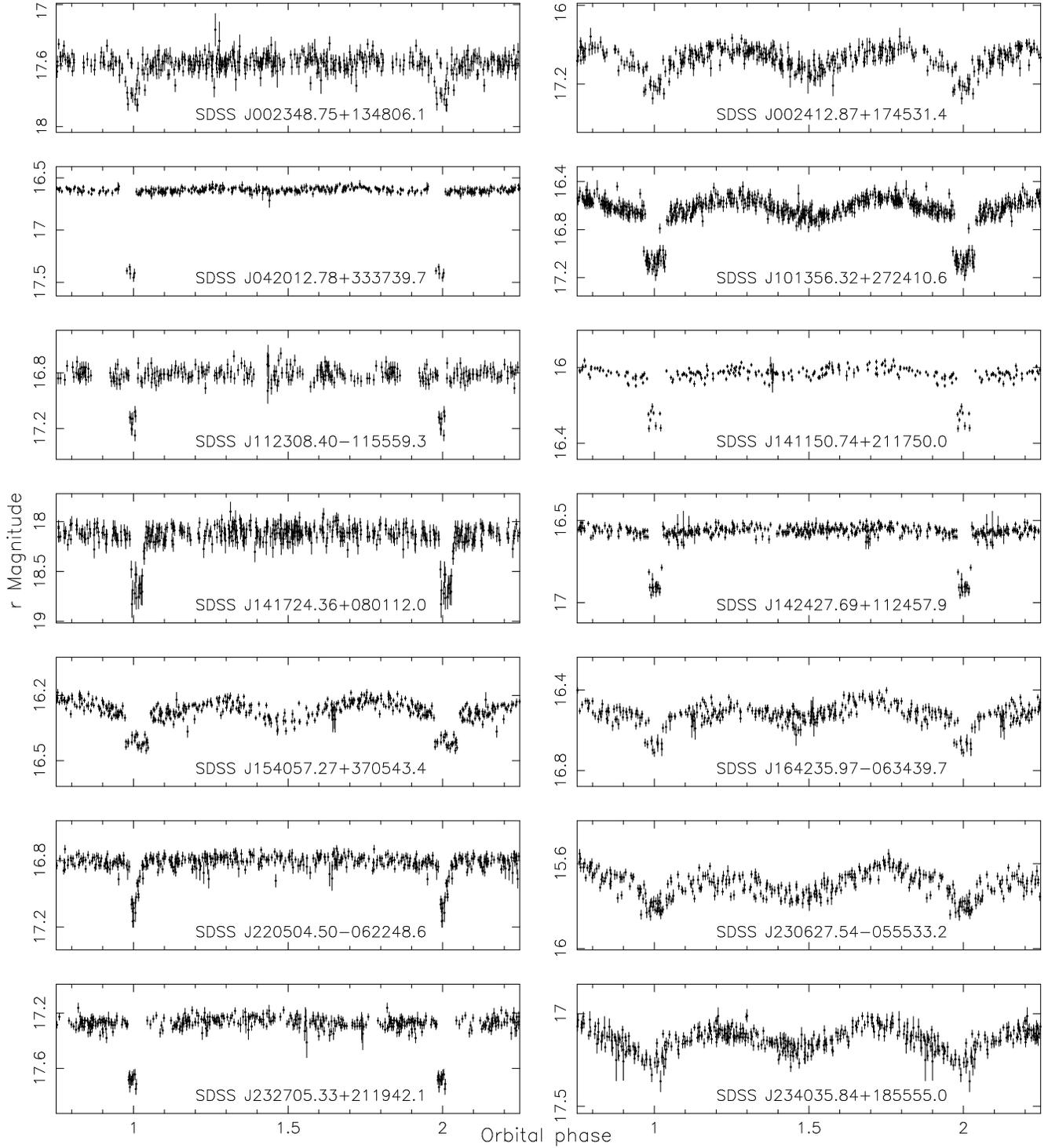}
 \vspace{20mm}
 \caption{Phase-folded CSS light curves of the 14 newly identified eclipsing
   PCEBs. SDSS\,J234035.84+185555.0 is a marginal detection and
     requires some follow-up data to confirm the eclipse.}
 \label{fig:crts_lcurves}
 \end{center}
\end{figure*}

\begin{figure*}
\begin{center}
 \includegraphics[width=\textwidth]{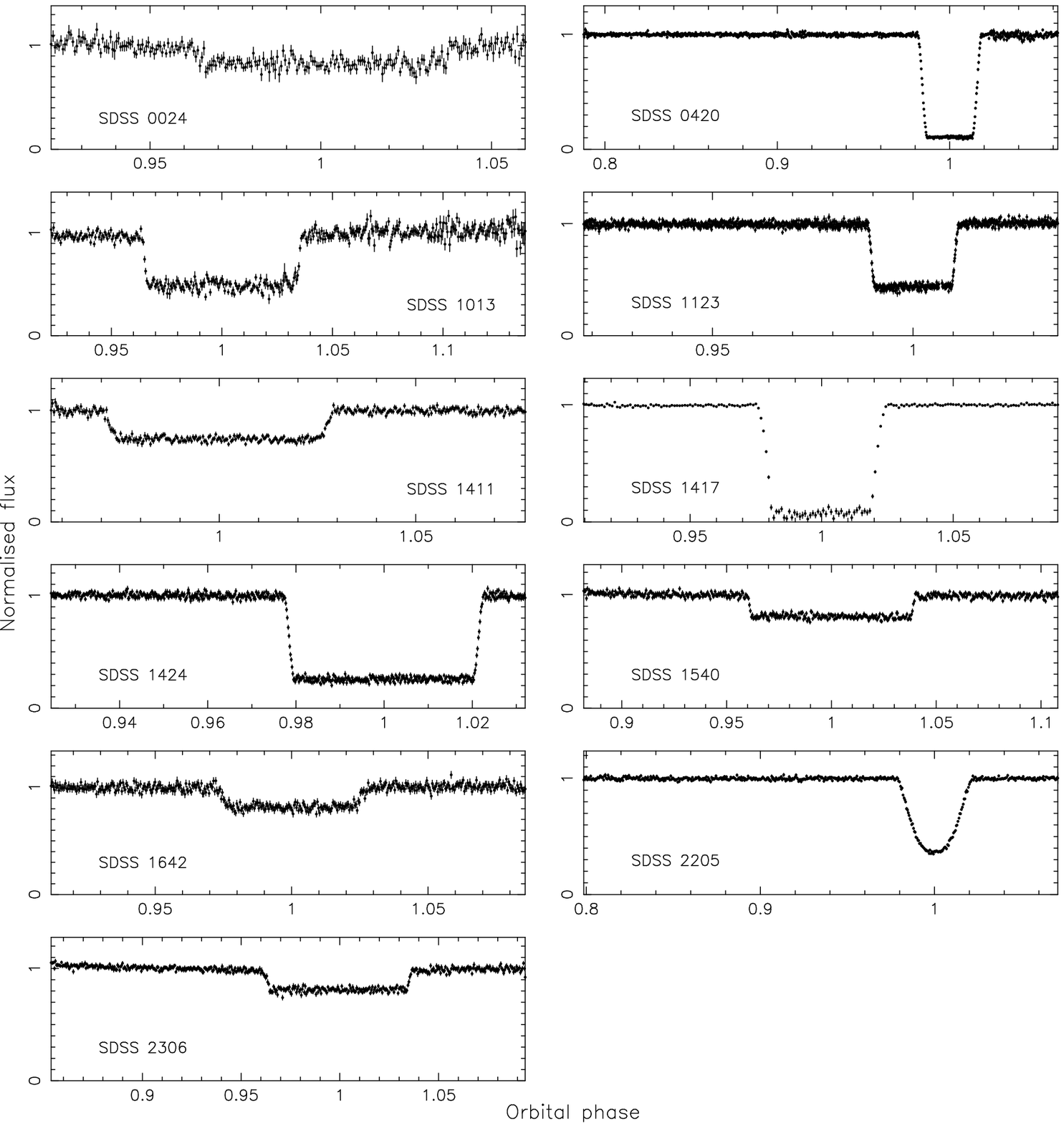}
 \caption{Follow-up light curves of 11 of the newly identified eclipsing
   systems, see Table~\ref{tab:photobslog} for the telescope+instrument+filter
   used for each observation. In the case of multiple bands we have
     plotted the $g'$ band eclipse.}
 \label{fig:followup}
 \end{center}
\end{figure*}

The spectra were substantially affected by fringing. This effect was
particularly pronounced at wavelengths beyond 8000{\AA}, therefore we only
used the reliable range of 5500{\AA}--8000{\AA} for determining the spectral
types of the main-sequence stars. Since the main-sequence stars dominate over
the white dwarfs in all our spectra we are unable to place any constraints on
the white dwarf parameters. We determined the spectral types of the
main-sequence stars using the technique outlined in \citet{rebassa07}, these
are listed in Table~\ref{tab:ellips} and Table~\ref{tab:app:eclipsers} with an
uncertainty of $\pm1$ class. An example fit is shown in
Figure~\ref{fig:specfit}. 

\section{Results}

We search for periodic signals in each of our 1958 CSS light curves using several
different period finding algorithms. We used a combination of Lomb-Scargle
\citep{scargle82} and an analysis of variance period search on the light curves
using a multi-harmonic model \citep{schwarzenberg96}, as well as phase
dispersion minimisation \citep{stellingwerf78}. These latter two approaches
are more sensitive to shallow eclipse features. We then used these results to
identify eclipsing systems, as well as any other periodically variable systems
(i.e. reflection effects and ellipsoidal variations). Finally, we also used a
simplified version of the Box-Fitting Least Squares method \citep{kovacs02},
described in more detail in \citet{parsons13css}

\begin{figure}
\begin{center}
 \includegraphics[width=\columnwidth]{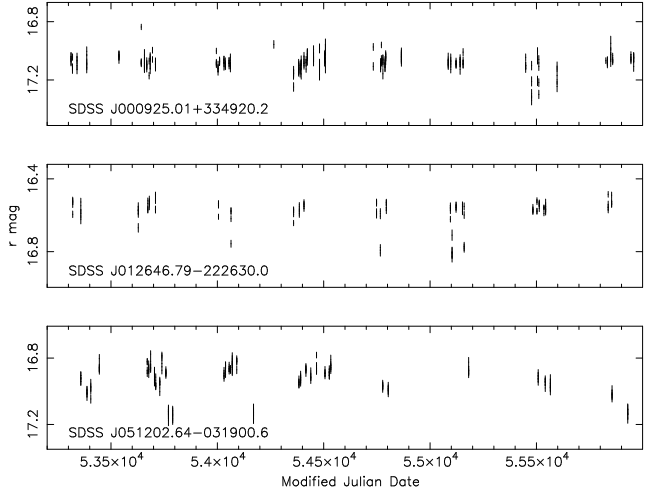}
  \caption{CSS light curves of the three candidate eclipsing WDMS systems. All
  show several fainter points indicative of an eclipse, but no obvious
  periodicity is found. More data are needed to determine their periods.}
 \label{fig:candidates}
 \end{center}
\end{figure}

\subsection{New eclipsing systems}

From our analysis we have identified a total of 17 eclipsing systems, 14
previously unknown systems and 3 systems that were already known to be
eclipsing PCEBs: WD\,1333+005 \citep{drake10ecl}, CSS\,J0314+0206
\citep{drake14pv} and CSS\,J0935+2700 \citep{drake14us}. For 11 of these 14
new systems we have obtained high-speed photometry of the eclipse of the white
dwarf. We list the details of these eclipsing systems, along with all other
known eclipsing white dwarf plus main-sequence / brown dwarf binaries (a total
of 71) in Table~\ref{tab:app:eclipsers}, in the appendix. The phase-folded CSS
light curves are shown in Figure~\ref{fig:crts_lcurves} and our follow-up
high-speed photometry of the white dwarf eclipses are shown in
Figure~\ref{fig:followup}. The majority of the new eclipsing systems show
shallow eclipses ($<$0.5\,mag) and large ellipsoidal modulation. We discuss
some of the more interesting systems in more detail in
Section~\ref{sec:indiv}. 

\begin{figure}
\begin{center}
 \includegraphics[width=\columnwidth]{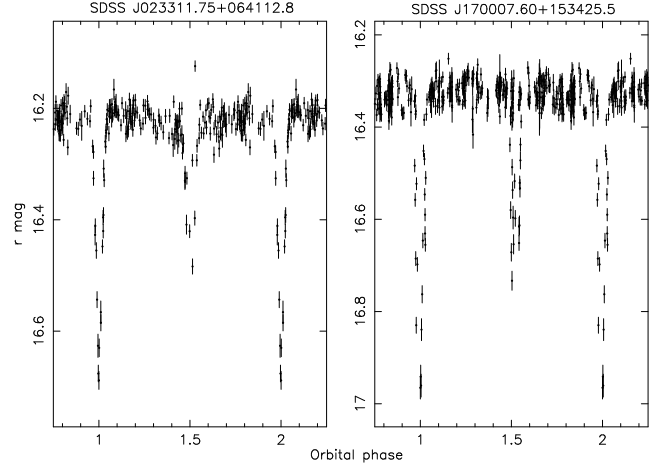}
  \caption{Phase-folded CSS light curves of the two newly identified eclipsing
    main-sequence plus main-sequence binaries.}
 \label{fig:msms}
 \end{center}
\end{figure}

Along with these new eclipsing systems, we identify three additional eclipsing
WDMS candidate systems. SDSS\,J000925.01+334920.2, SDSS\,J012646.79-222630.0
and SDSS\,J051202.64-031900.6 all show a number of faint points in their light
curves. However, none of them show any convincing periodic signals We show
their CSS light curves in Figure~\ref{fig:candidates}. We also
identify SDSS\,J023311.75+064112.8 and SDSS\,J170007.60+153425.5 as eclipsing
main-sequence plus main-sequence binaries, as revealed by the presence of deep
secondary eclipses, with periods of 0.5212d and 2.0482d respectively. We show
their phase-folded light curves in Figure~\ref{fig:msms}. 

\subsection{Non-eclipsing systems}

\begin{table}
 \centering
  \caption{Non-eclipsing variables showing ellipsoidal modulation (E) or
    reflection effects (R). We also list the amplitude of the variation and
    the spectroscopically measured spectral type of the main-sequence star in
    the binary (with an uncertainty of $\pm1$ class) for those with
    spectra. The systems highlighted in bold were previously discovered by
    \citet{drake14pv}.}
  \label{tab:ellips}
  \begin{tabular}{@{}lcccc@{}}
  \hline
  Object                    & Period     & Amp & Spectral & Type      \\
                            & (days)     & (mags)    & type     &           \\
  \hline
  SDSS\,J003159.53+263445.7 & 0.23857749 & 0.092     & M2       & E         \\
  {\bf SDSS\,J004638.02-033908.3} & 0.17490842 & 0.210     & M2       & E         \\
  SDSS\,J005400.33+142931.2 & 0.17178094 & 0.205     & M3       & E         \\
  SDSS\,J024924.76+071344.2 & 0.17333085 & 0.234     & M2       & E         \\
  {\bf SDSS\,J075015.11+494333.2} & 0.17300847 & 0.152     & -        & E         \\
  SDSS\,J081437.77+425236.6 & 0.21221388 & 0.046     & -        & E         \\
  {\bf SDSS\,J083405.20+300039.8} & 0.26143431 & 0.122     & -        & E         \\
  SDSS\,J083947.39+022332.0 & 0.10758061 & 0.117     & -        & E         \\
  {\bf SDSS\,J085414.28+211148.2} & 0.20430622 & 0.156     & -        & E         \\
  {\bf SDSS\,J090119.23+114255.1} & 0.18668802 & 0.256     & -        & E         \\
  SDSS\,J122942.09+105732.9 & 0.16602091 & 0.084     & -        & E         \\
  SDSS\,J135048.24+124630.8 & 0.81625256 & 0.068     & -        & E         \\
  {\bf SDSS\,J150606.71+215246.0} & 0.18594902 & 0.232     & M3       & E         \\
  SDSS\,J205321.72+001536.3 & 0.17897014 & 0.141     & M3       & E         \\
  SDSS\,J212244.75+103915.2 & 0.18130918 & 0.147     & M3       & E         \\
  SDSS\,J214857.39+192759.7 & 0.18455907 & 0.092     & M4       & R         \\
  {\bf SDSS\,J222918.95+185340.2} & 0.18918441 & 0.248     & M3       & E         \\
  {\bf SDSS\,J224134.90+253648.6} & 0.15737950 & 0.296     & M4       & E         \\
  SDSS\,J225428.64+131528.6 & 0.38236318 & 0.079     & -        & E         \\
  SDSS\,J225607.11+182245.6 & 0.16377587 & 0.206     & M3       & E         \\
  SDSS\,J233900.39+115707.2 & 0.12286761 & 0.148     & M5       & E         \\
  SDSS\,J235320.09+141532.5 & 0.15181691 & 0.120     & -        & E         \\
  \hline
  \end{tabular}
\end{table}

\begin{figure}
\begin{center}
 \includegraphics[width=0.98\columnwidth]{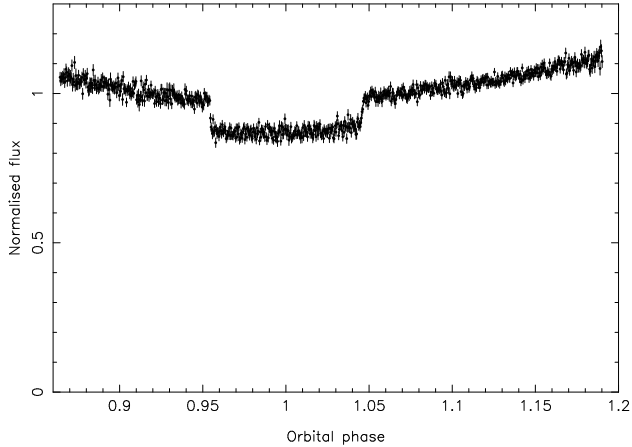}
 \caption{High-speed ULTRACAM $g'$ band light curve of the eclipsing binary
   SDSS\,J0745+2631. This system was classified as an eclipsing ``candidate''
   system by \citet{parsons13css} (it was not part of our photometric
     sample) because no eclipse could be detected in both 
   the CSS photometry and the LT follow-up photometry. Shorter wavelength
   photometry was required to confirm the eclipse, which was expected based on
   the large amplitude of the ellipsoidal modulation in the CSS light
   curve. Many of the systems listed in Table~\ref{tab:ellips} show very
   similar CSS light curves to this system and hence could also be eclipsing.}
 \label{fig:sdss0745_lc}
 \end{center}
\end{figure}

As well as eclipsing systems we also discovered 22 objects with periodically
varying light curves caused either by irradiation effects or 
the distorted shape of the main-sequence star, known as ellipsoidal
modulation. These are 
detailed in Table~\ref{tab:ellips}. Eight of these have 
previously been identified as close binaries by \citet{drake14pv,drake14us},
our measured periods are consistent with theirs. None of these light curves
showed any obvious eclipse features. However, as noted in
\citet{parsons13css}, systems that display large ellipsoidal amplitudes are
likely to also have high inclinations. Therefore, it is likely that some of
these may also be eclipsing, but the eclipse is too shallow to be detected
in the CSS photometry. This is demonstrated in Figure~\ref{fig:sdss0745_lc},
which shows a $g'$ band light curve of SDSS\,J0745+2631, a binary that displays
large ellipsoidal modulation (0.254 mags) in its CSS light
curve previously discovered in the SDSS spectroscopic sample by
\citet{parsons13css}. In that paper we were unable to confirm its eclipsing
nature because the eclipse is too shallow at the wavelengths of CSS and
the LT follow-up photometry that was obtained. However, as
Figure~\ref{fig:sdss0745_lc} shows, recently obtained ULTRACAM $g'$ band
shorter wavelength data reveals the presence of a shallow eclipse, confirming
that systems showing large ellipsoidal modulation are likely to have at least
shallow eclipses. 

\begin{figure*}
\begin{center}
 \includegraphics[width=0.98\columnwidth]{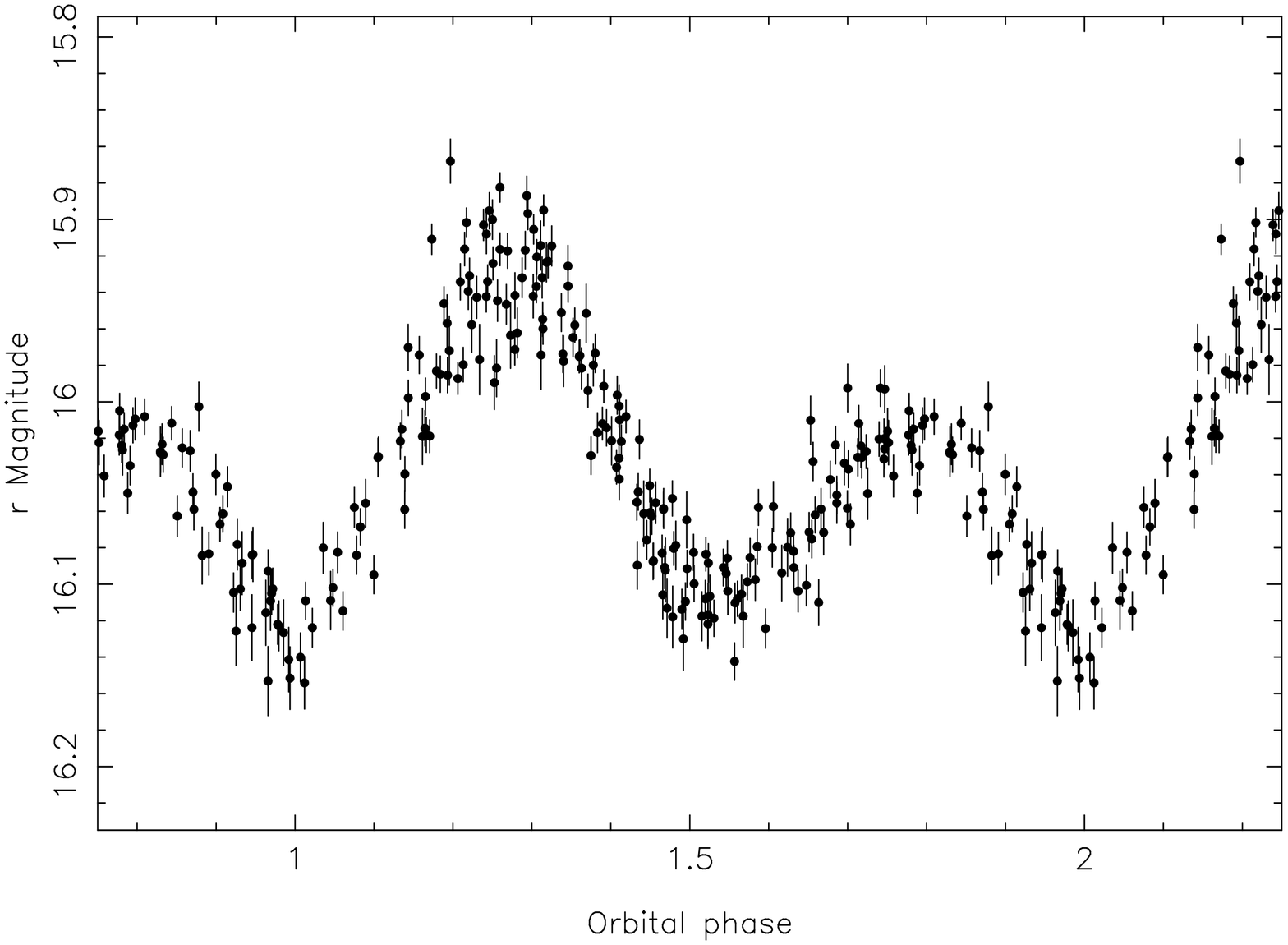}
\hspace{3mm}
 \includegraphics[width=0.95\columnwidth]{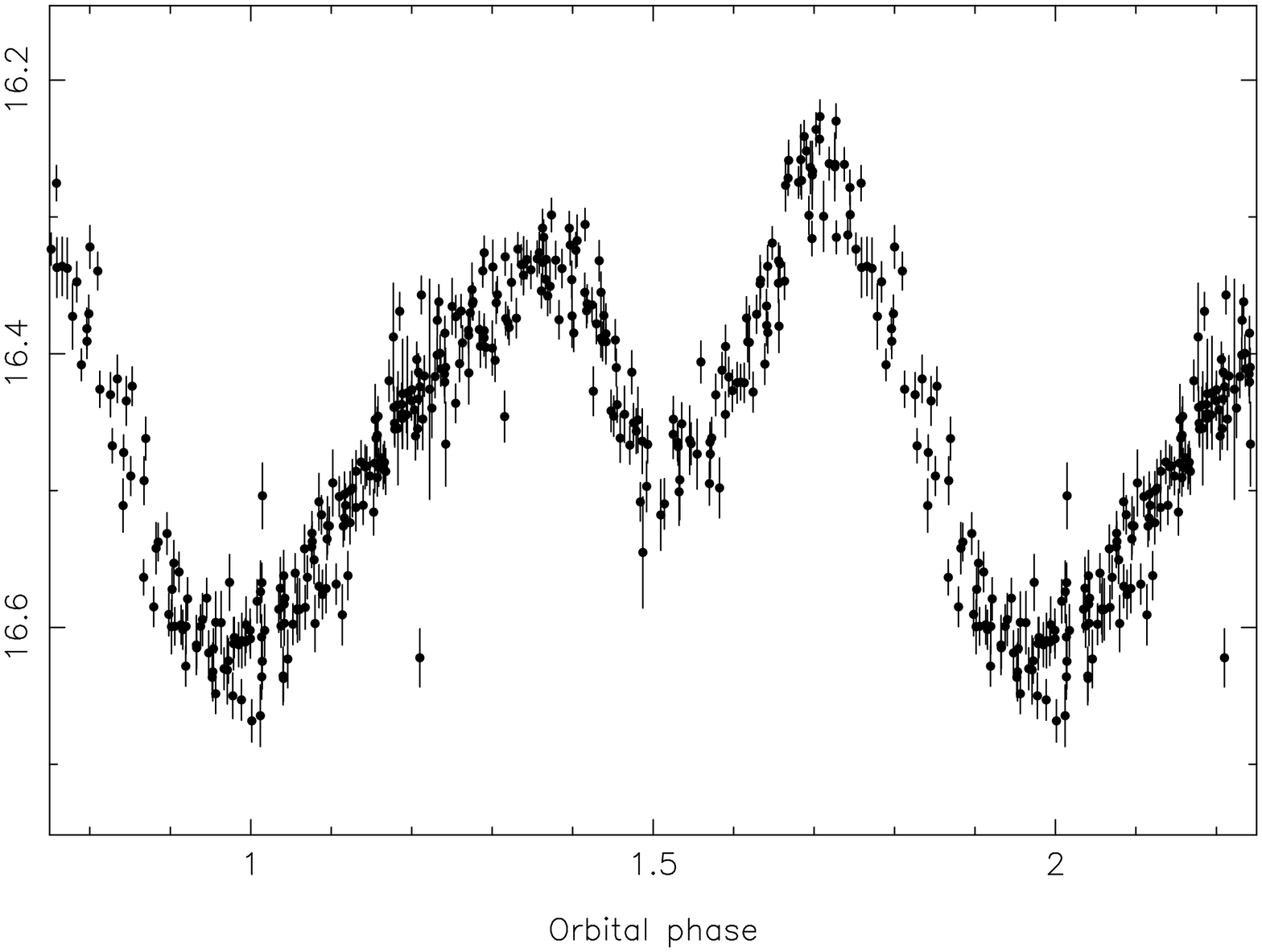}
  \caption{CSS light curves of SDSS\,J075015.11+494333.2 (left) and SDSS\,J222918.95+185340.2 (right). Both systems show variations similar to ellipsoidal modulation, but in both cases one maxima is lower than the other and for SDSS\,J222918.95+185340.2 the minima is substantially lower at phase 1.}
 \label{fig:odd_elip}
 \end{center}
\end{figure*}

Two of the close binaries listed in Table~\ref{tab:ellips} have slightly odd
light curves, which are shown in Figure~\ref{fig:odd_elip}. In both cases
there is a sinusoidal variation on half the orbital period, similar to
ellipsoidal modulation, but one peak is much stronger than the other, similar
to the O'Connell effect \citep{oconnell51}. This effect is often seen in PCEBs
with large Roche-lobe filling factors \citep{gansicke04,tappert07}, but is not
usually as stable on such long timescales. This could be accomplished by
having a large star spot on one side of the main-sequence star, but it would
have had to have persisted for the 8 years of CSS coverage, which is extremely
unlikely. Further data is needed to properly understand the nature of these
systems. The unusual light curve of SDSS\,J222918.95+185340.2 was previously
noted by \citet{drake14pv}, as an example of a difficult to classify light
curve. 

\section{Discussion} 

\subsection{Comparison with the spectroscopic sample}

\begin{figure}
\begin{center}
 \includegraphics[width=0.98\columnwidth]{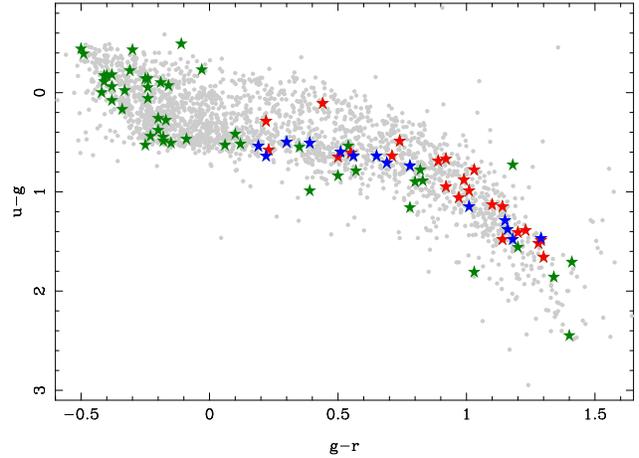}
 \caption{Distribution of all spectroscopic WDMS binaries (grey points) in the
   ($u-g,g-r$) colour plane. Previously discovered eclipsing PCEBs are shown
   in green and generally contain hotter white dwarfs and late-type
   main-sequence stars and are hence towards the bluer end of the
   distribution. The 14 newly discovered eclipsing PCEBs from this paper are
   shown in blue and those systems we found showing large ellipsoidal
   modulation are shown in red. Both these distributions have redder colours
   than the known eclipsing systems, indicative of their cooler white dwarfs
   and/or earlier spectral type main-sequence stars.}
 \label{fig:ccut}
 \end{center}
\end{figure}

Figure~\ref{fig:ccut} shows the colours of our newly discovered eclipsing and
ellipsoidal binaries compared to the SDSS spectroscopic sample
\citep{rebassa12}. Generally our newly discovered systems have much redder
colours than the eclipsing systems from the spectroscopic sample, indicating
that they contain cooler white dwarfs and/or earlier spectral type
main-sequence stars, as confirmed from our spectroscopic observations and as
expected from the colour selection itself \citep{rebassa13}. This is because
the colour cut was designed to carefully select WDMS systems whilst removing
as much contamination as possible. This required good 2MASS and WISE infrared
data (to remove quasars), which therefore selects against very late-type M
stars, since they are usually too faint to have good infrared photometry. 

For the spectroscopic WDMS sample $\sim$$1/3$ are thought to be close PCEBs
\citep{schreiber10,rebassa11}, of which $\sim$$10\%$ are eclipsing
\citep{parsons13css}. However, from our 1958 objects we only found 17
eclipsing systems, which is a much lower fraction than the spectroscopic
sample (we would have expected $\sim$$60$ eclipsing systems). Assuming that
the same fraction of close systems in our sample are eclipsing, this implies
that the close binary fraction in the colour selected sample is only
$\sim$$10\%$. Even if all 21 of the ellipsoidal systems are in fact eclipsing
(but not detectable in the CSS data), the fraction is still well below that of
the spectroscopic sample. 

There are three factors that contribute to this lower fraction. Firstly, as
evidenced by our detection of 2 eclipsing main-sequence plus main-sequence
binaries, the photometric sample is not pure, there is some
contamination. \citet{rebassa13} estimated that 84 per cent of the sample were
very likely genuine WDMS systems, which is consistent with our finding 2
non-WDMS eclipsing systems compared to 17 WDMS eclipsers. However, even taking
this contamination into account, the number of eclipsing systems is still well
below that of the spectroscopic sample. Secondly, as
previously noted, the colour selected sample contains systems with cooler
white dwarfs and/or early M-type main-sequence stars. Therefore, the eclipse
of the white dwarf is much shallower than the majority of those in the
spectroscopic sample and hence some eclipsing systems may have been missed due
to their eclipses not being detected in the CSS data (see the next
section). Finally, 
\citet{schreiber10} found that within the spectroscopic sample the close
binary fraction drops as one moves towards earlier type M stars. They
interpret this as evidence of different angular momentum loss rates for
different spectral types and therefore evidence of disrupted magnetic braking,
since binaries containing these partially convective stars loose angular
momentum much faster than their fully convective counterparts, and hence become
cataclysmic variables much faster and disappear from the sample. In fact,
\citet{schreiber10} found that the close binary fraction of WDMS systems with
main-sequence stars of M2--3 spectral type is $\sim$$10\%$, hence our results
are fully consistent with the spectroscopic sample, and further support the
idea of a spectral type dependent angular momentum loss rate. 

\subsection{Completeness} \label{sec:simu}

In order to better understand the low PCEB fraction of the photometric WDMS
sample compared to the spectroscopic one, we have simulated CRTS light curves
of the expected PCEB population within the photometric sample to see what
percentage of systems should be detectable from their CRTS light curves and
what parameters most affect this. 

This was achieved by using the $u-g$ colour of each object to predict the
temperature of the white dwarf and main-sequence star spectral type using the
constraints from Figure~4 of \citet{rebassa13}. We then used the mass-spectral
type relation of \citet{baraffe96} and mass-radius relationship of
\citet{morales10} to determine the radius of the main-sequence star, which we
consider to be the volume-averaged radius of the star.

We then randomly generated white dwarf masses and periods based on the
distributions found from the spectroscopic PCEB sample
\citep{zorotovic11,nebot11}. Random inclinations were also generated. Roche
distortion was then applied to the main-sequence stars and light curves
generated using a code specifically designed to simulate close binaries
containing white dwarfs (see \citealt{copperwheat10} for more details on the
light curve code). The light curves were sampled at the times corresponding to
the CRTS observations, and with the same photometric uncertainties. Finally,
we ran our period search algorithms on the resulting synthetic light curves and
considered a PCEB successfully detected if the correct period was returned (or
half the period, to account for ellipsoidal modulation systems).

The result of this simulation was that $\sim$$13.5\%$ of PCEBs should have been
detectable from their CRTS light curves, of these $\sim$$40\%$ are eclipsing.
The vast majority ($\sim$$95\%$) of the non-eclipsing systems were detected
from ellipsoidal variations, the rest were detected from reflection
effects, consistent with the results from Table~\ref{tab:ellips}.
Figure~\ref{fig:simul} shows the chance of detecting a PCEB (success rate) as
a function of the orbital period and main-sequence star mass. It is clear that
the closer to Roche-lobe filling the star is, the higher the detection
probability, since this leads to larger ellipsoidal modulation. The success
rate drops substantially at periods of more than 0.5 days, with essentially
only eclipsing systems being detected at these longer periods, consistent with
our results. 

If $13.5\%$ of PCEBs within the photometric sample are detectable from their
CRTS light curves, then our discovery of 38 PCEBs would imply a total number
of PCEBs in the photometric sample of $\sim$$280$, or $\sim$$14\%$ of the
sample. This increases to $\sim$$17\%$ if we take into account the expected
non-WDMS contamination. This is consistent with the results from the
spectroscopic sample, considering the earlier spectral types within the
photometric sample.

\subsection{The periods of the ellipsoidal modulation systems}

There is an interesting clustering of the periods in Table~\ref{tab:ellips},
where 11 of the 22 systems have periods between 4--4.5 hours (0.166--0.188
days) and spectral types of mostly M2--3. This is the expected period for
stars of this spectral type to fill their Roche-lobes \citep{baraffe98}, but
is slightly shorter than the periods of cataclysmic variables (CVs) containing
stars of this spectral type \citep{knigge11}, since they are driven slightly
out of thermal equilibrium by mass loss. Hence these systems are most likely
pre-CVs, right on the verge of starting mass transfer. 

\begin{figure}
\begin{center}
 \includegraphics[width=0.95\columnwidth]{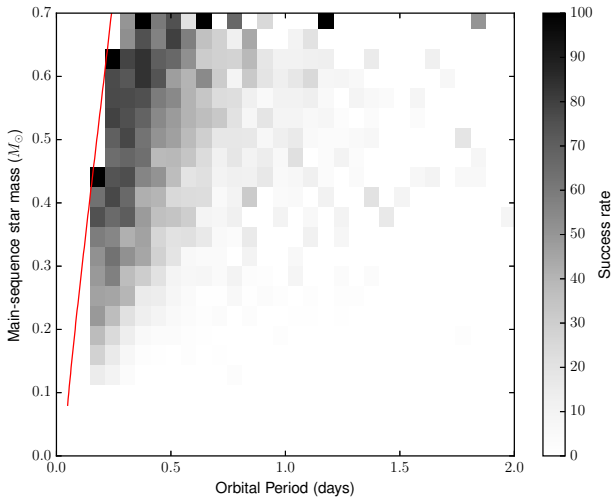}
 \caption{The chance of detecting a PCEB (success rate, see
   Section~\ref{sec:simu}) as a function of orbital period and main-sequence
   star mass. The red line shows where stars fill their
   Roche-lobes. Unsurprisingly, systems closer to Roche-lobe filling are more
   likely to be detected.}
 \label{fig:simul}
 \end{center}
\end{figure}

The high percentage of systems within this narrow period range ($50\%$ of the
ellipsoidal systems) is somewhat surprising. However, given that the
photometric sample appears to be dominated by main-sequence stars with M2--3
spectral types, and that ellipsoidal systems are easier to detect the closer
they are to Roche-lobe filling (see the previous section and
Figure~\ref{fig:simul}), this is the most likely period range in which to
detect these systems. At shorter periods they have become CVs, whilst at
longer periods the ellipsoidal amplitude is too low to detect within the CSS
photometry. 

\subsection{Notes on individual systems} \label{sec:indiv}

\subsubsection{SDSS\,J101356.32+272410.6}

This system is one of the handful of new DR 8 spectroscopic systems in our
sample and as such has an SDSS spectrum which has been decomposed and fitted
to determine the stellar parameters (see \citealt{rebassa07} for a detailed
explanation). The fit yields a massive white dwarf (1.14\MSUN,
\citealt{rebassa12}), which would be one of the most massive in an eclipsing
PCEB. However, as noted by \citet{parsons13css} the white dwarf parameters
determined from SDSS spectra may not be reliable if the spectrum is dominated
by the main-sequence star's features. Nevertheless, our high-speed eclipse
light curve (Figure~\ref{fig:followup}) shows that the ingress and egress are
very sharp, implying that the white dwarf is quite small and could therefore
be quite massive. 

\subsubsection{SDSS\,J112308.40-115559.3}

Like SDSS\,J101356.32+272410.6 this system has an SDSS spectrum which implies
that the white dwarf is also massive (1.1\MSUN, \citealt{rebassa12}). The
follow up photometry (Figure~\ref{fig:followup}) also revealed a fairly rapid
ingress and egress (hence small white dwarf) and so the high mass is
possible. However, this white dwarf is also fairly cool
($T_\mathrm{eff}=10073$\,K), which would make it small almost regardless of
the mass. Spectroscopic follow up is required to properly determine the
physical parameters of the system. 

\subsubsection{SDSS\,J220504.51-062248.4}

The spectroscopic fit to the main-sequence star in this system gives a robust
spectral type of M2. However, given the period of 3.1\,h we would expect such
a star to fill its Roche-lobe. There is no sign of this spectroscopically and
the CSS light curve shows very little out of eclipse variation, implying that
the system is not close to Roche-lobe filling. This also rules out an M2
subdwarf, since even with a higher density the system would still be very
close to Roche-lobe filling. This system was observed with ULTRACAM in
  the $u'$, $g'$ and $r'$ bands and showed that the object being eclipsed was
  bluer (hence hotter) than the M star, since the eclipse was deeper in the
  shorter wavelength bands. This, combined with the Galaxy Evolution Explorer
  (GALEX) magnitudes strongly imply that the eclipse is of a white dwarf by a
  low-mass star. However, We suspect that this system may in fact be a
triple system with a white dwarf in a close, eclipsing binary with a so-far
undetected low-mass star and with a wide M2 companion. Alternately, the
  M2 star may just lie along the line of sight to the binary.

\subsubsection{SDSS\,J234035.84+185555.0}

This system contains a main-sequence star of M1 spectral type, making it one
of the more massive main-sequence stars in an eclipsing PCEB. Only the K2 star
in V471 Tau \citep{obrien01} and the G5 star in KOI-3278 \citep{kruse14} are
more massive. Stars of this mass often show strong disagreement between their
measured physical parameters (e.g. radii, temperature) and theoretical
predictions \citep{lopez07} making this a useful system for investigating this
issue. 

\section{Conclusions}

We have combined a catalogue of photometrically selected white dwarf plus
main-sequence binaries from the Sloan Digital Sky Survey with the long-term
photometric data of the Catalina Sky Surveys in order to identify new close
post common-envelope binaries and specifically those that are eclipsing. We
have identified 17 eclipsing systems, of which 14 were previously unknown, as
well as 3 candidate eclipsing systems and 22 other close binaries. We have
presented follow-up photometry and spectroscopy for many of these in order to
better characterise their component stars.  

The newly discovered close binaries generally contain early M-type
main-sequence stars, that are underrepresented in the SDSS spectroscopic white
dwarf main-sequence star sample. These partially convective systems, in
combination with the spectroscopic white dwarf main-sequence star sample, will
offer valuable information on the angular momentum loss and evolution of close
binaries as well as constraining the mass-radius relationship of white dwarfs
and low-mass main-sequence stars.

\section*{Acknowledgments}

We thank the referee for useful comments and suggestions. SGP acknowledges
financial support from FONDECYT in the form of grant number 
3140585. ULTRACAM, VSD and SPL are supported by the Science and
Technology Facilities Council (STFC). TRM and EB were supported the STFC
\#ST/L000733/1. The research leading to these results
has received funding from the European Research Council under the European
Union's Seventh Framework Programme (FP/2007-2013) / ERC Grant Agreement
n. 320964 (WDTracer). MRS thanks for support from FONDECYT (1141269) and
Millennium Science Initiative, Chilean ministry of Economy: Nucleus
RC130007. ARM acknowledges   financial  support   from   the  Postdoctoral
Science Foundation of China  (grants 2013M530470 and 2014T70010)  and from the
Research  Fund  for International  Young  Scientists  by the  National Natural
Science   Foundation  of   China  (grant   11350110496). We also acknowledge
the travel support provided by the Royal Society. This work has made use of
data obtained at the Thai National Observatory on Doi Inthanon, operated by
NARIT.

\bibliographystyle{mn_new}
\bibliography{crts2}

\appendix

\section{Eclipsing white dwarf plus main-sequence / brown dwarf binaries}
\begin{table*}
 \centering
  \caption{Detached, eclipsing white dwarf plus main-sequence / brown dwarf
    binaries. Our newly discovered systems are highlighted in bold. See
    \citet{marsh14} for an explanation of the timing system used for T0, the
    mid-eclipse time. References: (1) This paper, (2) \citet{law12}, (3) \citet{kleinman04}, (4) \citet{pyrzas09}, (5) \citet{parsons12uc}, (6) \citet{becker11}, (7) \citet{parsons13mag}, (8) \citet{drake14pv}, (9) \citet{obrien01}, (10) \citet{maxted07}, (11) \citet{parsons13css}, (12) \citet{drake10ecl}, (13) \citet{parsons12lm}, (14) \citet{drake14us}, (15) \citet{pyrzas12}, (16) \citet{parsons12gs}, (17) \citet{vdbess07}, (18) \citet{odon03}, (19) \citet{littlefair14}, (20) \citet{parsons10nn}, (21) \citet{muirhead13}, (22) \citet{kruse14}, (23) \citet{almenara12}, (24) \citet{maxted04}.}
  \label{tab:app:eclipsers}
  \begin{tabular}{@{}lccccccccc@{}}
\hline
Name                   & RA          & Dec         & Period          & T0              & WD mass & WD $T_\mathrm{eff}$ & MS star  & $r$ mag & Ref \\
                       &             &             & (Days)          & BMJD(TDB)       & (\MSUN) & (K)               & sp type  &         &     \\
\hline
{\bf SDSS\,J0023+1348} & 00:23:48.75 & +13:48:06.1 & 0.353930(5)     & 54733.2550(10)  &         &                   &          & 18.2    & 1   \\
{\bf SDSS\,J0024+1745} & 00:24:12.87 & +17:45:31.4 & 0.2000379(5)    & 56482.19681(9)  &         &                   &          & 17.6    & 1   \\
PTFEB11.441            & 00:45:46.00 & +41:50:30.0 & 0.35871(5)      & 55437.8165(1)   & 0.54    & 8500              & M3       & 16.4    & 2   \\
SDSS\,J0106-0014       & 01:06:22.99 & -00:14:56.2 & 0.08501533140(2)& 55059.056102(2) & 0.37    & 14393             & M8       & 18.4    & 3   \\
SDSS\,J0110+1326       & 01:10:09.09 & +13:26:16.1 & 0.3326867540(2) & 53993.949088(1) & 0.47    & 25900             & M4       & 16.9    & 4   \\
SDSS\,J0138-0016       & 01:38:51.54 & -00:16:21.6 & 0.072764956(5)  & 55867.007405(7) & 0.529   & 3570              & M5       & 17.4    & 5   \\
PTFEB28.235            & 01:52:56.60 & +38:44:13.4 & 0.38612033(3)   & 55459.60923(7)  & 0.6     & 8000              & M3       & 16.6    & 2   \\
PTFEB28.852            & 01:55:24.70 & +37:31:53.8 & 0.46152(9)      & 55530.2335(1)   & 0.49    & 8500              & M2       & 17.2    & 2   \\
SDSS\,J0259-0044       & 02:59:53.33 & -00:44:00.2 & 0.144183554(33) & 51819.4150(10)  &         &                   & M3       & 19.3    & 6   \\
SDSS\,J0303+0054       & 03:03:08.35 & +00:54:44.1 & 0.1344376668(1) & 53991.117307(2) & 0.839   & 8000              & M4.5     & 18.1    & 7   \\
SDSS\,J0308-0054       & 03:08:56.55 & -00:54:50.7 & 0.18595942(1)   & 56181.14358(3)  &         &                   & M4.5     & 17.4    & 6   \\
CSS\,J0314+0206        & 03:14:52.11 & +02:06:07.1 & 0.305296755(4)  & 56195.206350(2) &         &                   &          & 17.4    & 8   \\
CSS\,J0344+0930        & 03:44:20.14 & +09:30:07.1 & 0.28066299(1)   & 56328.1540(5)   &         &                   &          & 18.2    & 8   \\
V471\,Tau              & 03:50:24.97 & +17:14:47.4 & 0.52118343117(8)& 54027.953026(1) & 0.84    & 34500             & K2       &  9.4    & 9   \\
{\bf SDSS\,J0420+3337} & 04:20:12.78 & +33:37:39.7 & 0.2205602(5)    & 56721.890778(2) &         &                   &          & 16.9    & 1   \\
RR\,Cae                & 04:21:05.53 & -48:39:08.3 & 0.30370368162(5)& 51522.5484387(5)& 0.44    & 7540              & M5       & 14.3    & 10  \\
SDSS\,J0745+2631       & 07:45:48.63 & +26:31:23.4 & 0.21926417(8)   & 53387.2495(10)  &         &                   & M2       & 17.4    & 11  \\
SDSS\,J0821+4559       & 08:21:45.27 & +45:59:23.4 & 0.50909205(5)   & 55989.038796(23)& 0.66    & 80938             & M2       & 17.6    & 11  \\
CSS\,40190             & 08:38:45.86 & +19:14:16.5 & 0.130112321(1)  & 53469.21993(3)  & 0.5     & 11062             & M7       & 18.4    & 12  \\
SDSS\,J0857+0342       & 08:57:46.18 & +03:42:55.3 & 0.0650965381(2) & 55552.7127649(7)& 0.514   & 37400             & M8       & 18.3    & 13  \\
CSS\,080502            & 09:08:12.04 & +06:04:21.2 & 0.1494380537(5) & 53466.33317(4)  & 0.35    & 17505             & M4       & 17.3    & 12  \\
LP\,486-53             & 09:08:26.18 & +12:36:48.9 & 0.13919939(1)   & 54140.3160(5)   &         &                   &          & 16.5    & 14  \\
SDSS\,J0927+3329       & 09:27:41.73 & +33:29:59.1 & 2.308227(1)     & 56074.906137(21)& 0.59    & 27111             & M3       & 18.2    & 11  \\
CSS\,J0935+2700        & 09:35:08.00 & +27:00:49.2 & 0.20103360(2)   & 56602.839467(8) &         &                   &          & 16.6    & 14  \\
CSS\,38094             & 09:39:47.95 & +32:58;07.3 & 0.330989669(3)  & 55587.308818(8) & 0.49    & 28442             & M5       & 18.0    & 12  \\
SDSS\,J0946+2030       & 09:46:34.49 & +20:30:03.4 & 0.25286147(5)   & 56032.94559(3)  & 0.62    & 10307             & M5       & 18.9    & 11  \\
CSS\,41631             & 09:57:19.24 & +23:42:40.7 & 0.150870774(1)  & 55548.357087(2) & 0.43    & 25891             & M2       & 18.1    & 12  \\
SDSS\,J0957+3001       & 09:57:37.59 & +30:01:36.5 & 1.926125(1)     & 56014.975114(32)& 0.42    & 28064             & M3       & 18.8    & 11  \\
{\bf SDSS\,J1013+2724} & 10:13:56.32 & +27:24:10.6 & 0.129040379(4)  & 56280.182901(8) & 1.14    & 15601             & M4       & 17.2    & 1   \\
SDSS\,J1021+1744       & 10:21:02.25 & +17:44:39.9 & 0.14035862(3)   & 56093.9055(1)   & 0.5     & 32595             & M4       & 19.0    & 11  \\
SDSS\,J1028+0931       & 10:28:57.78 & +09:31:29.8 & 0.235025695(6)  & 56001.093511(9) & 0.42    & 18756             & M3       & 15.6    & 11  \\
SDSS\,J1057+1307       & 10:57:56.93 & +13:07:03.5 & 0.125162238(3)  & 56010.062214(14)& 0.34    & 12536             & M5       & 18.6    & 11  \\
{\bf SDSS\,J1123-1155} & 11:23:08.40 & -11:55:59.3 & 0.7691358(14)   & 56685.789599(9) & 1.26    & 10073             & M5       & 17.5    & 1   \\
SDSS\,J1210+3347       & 12:10:10.13 & +33:47:22.9 & 0.124489764(1)  & 54923.033686(6) & 0.415   & 6000              & M5       & 16.2    & 15  \\
SDSS\,J1212-0123       & 12:12:58.25 & -01:23:10.2 & 0.335870942(4)  & 54104.20914(2)  & 0.439   & 17900             & M4       & 16.9    & 16  \\
SDSS\,J1223-0056       & 12:23:39.61 & -00:56:31.1 & 0.0900784(13)   & 55707.016987(7) & 0.45    & 11565             & M6       & 18.0    & 11  \\
CSS\,25601             & 12:44:32.25 & +10:17:10.8 & 0.22785631(9)   & 53466.3611(11)  & 0.37    & 21168             & M5       & 18.3    & 12  \\
SDSS\,J1307+2156       & 13:07:33.49 & +21:56:36.7 & 0.21632224(1)   & 56007.22137(2)  &         &                   & M4       & 17.4    & 11  \\
CSS\,21616             & 13:25:18.18 & +23:38:07.9 & 0.194958992(2)  & 55653.454186(9) &         &                   &          & 18.3    & 12  \\
DE\,CVn                & 13:26:53.26 & +45:32:46.7 & 0.3641393156(5) & 52784.054043(1) & 0.51    & 8000              & M3       & 12.5    & 17  \\
SDSS\,J1329+1230       & 13:29:25.21 & +12:30:25.4 & 0.080966255(1)  & 55271.054818(1) & 0.35    & 13000             & M8       & 17.5    & 11  \\
WD\,1333+005           & 13:36:16.05 & +00:17:31.9 & 0.1219587642(7) & 55611.47667(2)  &         & 6000              & M5       & 17.1    & 12  \\
CSS\,21357             & 13:48:41.61 & +18:34:10.5 & 0.248431787(2)  & 56000.161910(7) & 0.59    & 15071             & M4       & 17.2    & 12  \\
QS\,Vir                & 13:49:51.95 & -13:13:37.5 & 0.150757612(1)  & 48689.13553(5)  & 0.77    & 14085             & M3       & 14.1    & 18  \\
SDSS\,J1408+2950       & 14:08:47.14 & +29:50:44.9 & 0.19179048(4)   & 53506.2885(5)   & 0.49    & 29050             & M5       & 19.0    & 11  \\
CSS\,07125             & 14:10:57.73 & -02:02:36.7 & 0.36349709(2)   & 53464.48862(24) & 0.38    & 29268             & M4       & 18.9    & 12  \\
CSS\,21055             & 14:11:26.20 & +20:09:11.1 & 0.0845327526(13)& 55991.388717(2) & 0.53    & 13000             & L7       & 18.0    & 19  \\
SDSS\,J1411+1028       & 14:11:34.70 & +10:28:39.7 & 0.16750961(3)   & 56031.17278(5)  & 0.36    & 30419             & M3       & 19.2    & 11  \\
{\bf SDSS\,J1411+2117} & 14:11:50.74 & +21:17:50.0 & 0.3216383(1)    & 56396.117096(24)&         &                   & M3       & 16.4    & 1   \\
GK\,Vir                & 14:15:36.41 & +01:17:18.2 & 0.3443308387(1) & 42543.337714(3) & 0.563   & 55000             & M4.5     & 17.3    & 16  \\
{\bf SDSS\,J1417+0801} & 14:17:24.36 & +08:01:12.0 & 0.23918209(5)   & 56826.97958(1)  &         &                   &          & 18.6    & 1   \\
CSS\,080408            & 14:23:55.06 & +24:09:24.3 & 0.382004302(6)  & 53470.39959(4)  & 0.4     & 32595             & M5       & 17.9    & 12  \\
{\bf SDSS\,J1424+1124} & 14:24:27.69 & +11:24:57.9 & 0.2392937(1)    & 56391.123610(8) &         &                   &          & 17.1    & 1   \\
SDSS\,J1435+3733       & 14:35:47.87 & +37:33:38.5 & 0.1256310520(3) & 54148.204447(3) & 0.505   & 12500             & M5       & 17.3    & 4   \\
CSS\,09797             & 14:56:34.30 & +16:11:37.7 & 0.2291202(2)    & 51665.7893(30)  & 0.34    & 19149             & M5       & 18.0    & 12  \\
{\bf SDSS\,J1540+3705} & 15:40:57.27 & +37:05:43.4 & 0.26143467(5)   & 56434.18457(3)  &         &                   & M4       & 16.8    & 1   \\
\hline
\end{tabular}
\end{table*}

\begin{table*}
\centering
\begin{tabular}{@{}lccccccccc@{}}
\hline
Name                   & RA          & Dec         & Period          & T0              & WD mass & WD $T_\mathrm{eff}$ & MS star  & $r$ mag & Ref \\
                       &             &             & (Days)          & BMJD(TDB)       & (\MSUN) & (K)               & sp type  &         &     \\
\hline
SDSS\,J1548+4057       & 15:48:46.00 & +40:57:28.8 & 0.185515281(1)  & 54592.07295(3)  & 0.646   & 11700             & M6       & 18.4    & 4   \\
NN\,Ser                & 15:52:56.20 & +12:54:47.2 & 0.1300801218(1) & 47344.0254693(7)& 0.535   & 57000             & M4       & 16.9    & 20  \\
{\bf SDSS\,J1642-0634} & 16:42:35.97 & -06:34:39.7 & 0.28689(1)      & 56770.19243(3)  &         &                   &          & 17.1    & 1   \\
CSS\,J1653+3914        & 16:53:52.40 & +39:14:10.3 & 0.21981035(1)   & 54919.4450(5)   &         &                   &          & 16.5    & 14  \\
KOI-256                & 19:00:44.43 & +49:33:55.3 & 1.3786548(1)    & 55373.824838(1) & 0.592   & 7100              & M3       & 15.8    & 21  \\
KOI-3278               & 19:26:05.97 & +38:27:21.1 & 88.18052(27)    & 55084.9190(23)  & 0.634   & 9960              & G5       & 14.6    & 22  \\
2MASS J1942+4745       & 19:42:37.20 & +47:45:48.5 & 0.350468722(6)  & 53589.93613(1)  & 0.61    & 20470             & M4       & 18.7    & 23  \\
RX\,J2130.6+4710       & 21:30:18.60 & +47:10:08.0 & 0.5210361666(1) & 52785.1827943(5)& 0.554   & 18000             & M3.5     & 14.2    & 24  \\
CSS\,J2131+0758        & 21:31:02.61 & +07:58:10.1 & 0.24789997(5)   & 54730.2735(5)   &         &                   &          & 18.5    & 8   \\
{\bf SDSS\,J2205-0622} & 22:05:04.51 & -06:22:48.4 & 0.132386898(4)  & 56178.079537(3) &         &                   & M2       & 17.1    & 1   \\
CSS\,09704             & 22:08:23.66 & -01:15:34.1 & 0.156505698(1)  & 53507.45739(2)  & 0.37    & 29302             & M4       & 18.8    & 12  \\
SDSS\,J2235+1428       & 22:35:30.61 & +14:28:55.1 & 0.144456765(1)  & 55469.065472(6) & 0.45    & 21045             & M4       & 18.8    & 11  \\
{\bf SDSS\,J2306-0555} & 23:06:27.54 & -05:55:33.2 & 0.20008319(6)   & 57024.53910(2)  &         &                   &          & 16.2    & 1   \\
{\bf SDSS\,J2327+2119} & 23:27:05.33 & +21:19:42.1 & 0.84974077(5)   & 54363.2180(10)  &         &                   & M4       & 17.8    & 1   \\
{\bf SDSS\,J2340+1855} & 23:40:35.84 & +18:55:55.0 & 0.30765754(5)   & 54977.4535(10)  &         &                   & M1       & 17.2    & 1   \\
\hline
\end{tabular}
\end{table*}

\label{lastpage}

\end{document}